\documentclass{article}
\usepackage{ismir,amsmath,cite,url}
\usepackage{graphicx}
\usepackage{color}

\usepackage{mathtools}
\DeclarePairedDelimiter\abs{\lvert}{\rvert}%

\let\oldfootnote\footnote
\def\footnote{\ifhmode\unskip\fi\oldfootnote}

\title{An energy-based generative sequence model for testing sensory theories of Western harmony}

\twoauthors
 {Peter M. C. Harrison} {Queen Mary University of London \\ Cognitive Science Research Group}
 {Marcus T. Pearce} {Queen Mary University of London \\ Cognitive Science Research Group}

\begin{document}

\maketitle

\begin{abstract}
The relationship between sensory consonance and Western harmony
is an important topic in music theory and psychology.
We introduce new methods
for analysing this relationship,
and apply them to
large corpora representing three prominent genres of Western music:
classical, popular, and jazz music.
These methods centre on a generative sequence model
with an exponential-family energy-based form
that predicts chord sequences from
continuous features.
We use this model to investigate one aspect of
instantaneous consonance (harmonicity)
and two aspects of sequential consonance
(spectral distance and voice-leading distance).
Applied to our three musical genres,
the results generally support the relationship
between sensory consonance and harmony,
but lead us to question the high importance attributed to spectral distance
in the psychological literature.
We anticipate that our methods will provide a useful platform
for future work linking music psychology to music theory.
\end{abstract}

\section{Introduction}\label{sec:introduction}
Music theorists and psychologists have long sought to understand 
how Western harmony may be shaped by natural phenomena
universal to all humans
\cite{Helmholtz1954,Parncutt1989,Stumpf2012}. 
Key to this work is the notion of \textit{sensory consonance},
describing a sound's natural pleasantness
\cite{Regnault2001,Schellenberg1996a,Trainor2002},
and its inverse \textit{sensory dissonance},
describing natural unpleasantness.

Sensory consonance has both 
\textit{instantaneous} and \textit{sequential} aspects.
Instantaneous consonance is the consonance of an individual sound,
whereas sequential consonance is a property of 
a progression between sounds.

Instantaneous sensory consonance primarily derives
from \textit{roughness} and \textit{harmonicity}.
Roughness is an unpleasant sensation caused by 
interactions between spectral components in the inner ear
\cite{Daniel1997, Vencovsky2016},
whereas harmonicity 
\footnote{
Related concepts include
\textit{tonalness} \cite{Parncutt1989},
\textit{toneness} \cite{Huron2001},
\textit{fusion} \cite{Huron1991,Stumpf2012},
\textit{complex sonorousness} \cite{Parncutt1994},
and \textit{multiplicity} \cite{Parncutt1994}.
}
is a pleasant percept elicited by a sound's
resemblance to the harmonic series
\cite{McDermott2010, Bowling2015}.

Sequential sensory consonance is primarily determined by
\textit{spectral distance}
and \textit{voice-leading distance}.
Spectral distance
\footnote{
Spectral distance is also known by its antonym
\textit{spectral similarity} \cite{Milne2015}.
\textit{Pitch commonality} \cite{Parncutt1994} is a similar concept.
Psychological models of harmony and tonality in the 
auditory short-term memory (ASTM) tradition
typically rely on some kind of spectral distance measure
\cite{Bigand2014,Collins2014,Leman2000}.
}
describes how much
a sound's acoustic spectrum perceptually differs
from neighbouring spectra
\cite{Milne2011,Milne2015,Milne2016a,Parncutt1989,Parncutt1994}.
Voice-leading distance
\footnote{
Voice-leading distance is termed \textit{horizontal motion}
in \cite{Bigand1996}.
Parncutt's notion of \textit{pitch distance} \cite{Parncutt1994,Parncutt2011a} is also conceptually similar 
to voice-leading distance.
}
describes how far notes in one chord
have to move to produce the next chord
\cite{Bigand1996,Tymoczko2006,Tymoczko2011}.
Consonance is associated with low
spectral and voice-leading distance.

Many Western harmonic conventions can be rationalized as attempts to increase pleasantness by maximizing sensory consonance.
The major triad maximizes consonance by minimizing roughness and maximizing harmonicity;
the circle of fifths maximizes consonance by minimizing spectral distance;
tritone substitutions are consonant through voice-leading efficiency \cite{Tymoczko2006}.

This idea -- that Western harmony seeks to maximize sensory consonance -- has a long history in music theory \cite{Rameau1722}.
Its empirical support is surprisingly limited, however.
The best evidence comes from research linking sensory consonance maximization to rules from music theory \cite{Parncutt1989,Huron2001,Tymoczko2006},
but this work is constrained by the subjectivity and limited scope of music-theoretic textbooks.

A better approach is to bypass textbooks and analyse musical scores directly.
Usefully, large datasets of digitised musical scores are now available,
as are many computational models of consonance.
However, statistically linking them is non-trivial.
One could calculate distributions of consonance features, but this would give only limited causal insight into how these distributions arise.
Better insight might be achieved by regressing transition probabilities against consonance features,
but this approach is statistically problematic 
because of variance heterogeneity induced by the inevitable sparsity of the transition tables.

This paper presents a new statistical model developed for tackling this problem.
The model is generative and feature-based,
defining a probability distribution over symbolic sequences based on features derived from these sequences.
Unlike previous feature-based sequence models, it is specialized for continuous features,
making it well-suited to consonance modelling.
Moreover, the model parameters are easily interpretable and have quantifiable uncertainty, enabling error-controlled statistical inference.

We use this new model to test sensory theories of harmony as follows.
We fit the model to corpora of chord sequences from classical, popular, and jazz music, using psychological models of sensory consonance as features.
We then compute feature importance metrics to quantify how different aspects of consonance constrain harmonic movement.
This work constitutes the first corpus analysis comprehensively linking sensory consonance to harmonic practice.
\section{Methods}\label{sec:methods}

\subsection{Representations}

\subsubsection{Input}

Chord progressions are represented as sequences of pitch-class sets.
Exact chord repetitions are removed,
but changes of chord inversion are represented as 
repeated pitch-class sets.



\subsubsection{Pitch-Class Spectra} \label{pitch_class_spectra}

Some of our features use \textit{pitch-class spectra} as defined in \cite{Milne2011,Milne2016a}.
A pitch-class spectrum is a continuous function that describes
\textit{perceptual weight} as a function of pitch class ($p_c$).
Perceptual weight is the strength of perceptual
evidence for a given pitch class being present.
Pitch classes ($p_c$) take values in the interval $[0, 12)$
and relate to frequency ($f$, Hz scale)
as follows:


\begin{equation}
p_c = \left[9 + 12 \log_2{\left(\frac{f}{440}\right)}\right] \bmod{12}.
\end{equation}




Pitch-class sets are transformed to pitch-class spectra by
expanding each pitch class into its implied harmonics.
Pitch classes are modelled as harmonic complex tones with 12 harmonics, after \cite{Milne2016a}.
The $j$th harmonic in a pitch class
has level $j^{-\rho}$,
where $\rho$ is the roll-off parameter ($\rho > 0$).
Partials are represented by Gaussians
with mass equal to partial level,
mean equal to partial pitch class,
and standard deviation $\sigma$.
Perceptual weights combine additively.

Formally, $W(p_c, X)$ defines a pitch-class spectrum,
returning the perceptual weight at pitch-class $p_c$
for an input pitch-class set
$X = \left\{ x_1, x_2, \ldots, x_{m} \right\}$:

\begin{equation} \label{eq:pc_spec}
W(p_c, X) =
\sum_{i=1}^{m}
T(p_c, x_i).
\end{equation}

\noindent Here $i$ indexes the pitch classes, and
$T(p_c, x)$ is the contribution of a
harmonic complex tone with fundamental pitch class $x$
to an observation at pitch class $p_c$:

\begin{equation}
T(p_c, x) =
\sum_{j=1}^{12}
g\left(p_c, j^{-\rho}, h(x, j)\right).
\end{equation}

\noindent Now $j$ indexes the harmonics,
$g(p_c, l, p_x)$ is the contribution
from a harmonic with level $l$ and pitch-class $p_x$
to an observation at pitch-class $p_c$,

\begin{equation}
g(p_c, l, p_x) =
\frac{l}{\sigma\sqrt{2\pi}}
\exp{\left(- \frac{1}{2} \left(\frac{d(p_c, p_x)}{\sigma}\right)^2\right)},
\end{equation}

\noindent $d(p_x, p_y)$ is the distance between two pitch classes
$p_x$ and $p_y$,

\begin{equation} \label{eq:pitch_class_distance}
d(p_x, p_y) = 
\min \left(
\abs*{p_x - p_y},
12 - \abs*{p_x - p_y}
\right),
\end{equation}

\noindent
and $h(x, j)$ is the pitch class of the $j$th partial
of a harmonic complex tone with fundamental pitch class $x$:

\begin{equation}
h(x, j) = \left(x + 12 \log_2{j}\right) \bmod 12.
\end{equation}


\noindent $\rho$ and $\sigma$ are set to
0.75 and 0.0683 after \cite{Milne2016a}.

\subsection{Features}

\subsubsection{Spectral Distance}

\textit{Spectral distance} is operationalised using the psychological model of \cite{Milne2011,Milne2016a}.
The spectral distance between two pitch-class sets
$X, Y$ is defined as
1 minus
the continuous cosine similarity between the two pitch-class spectra:

\begin{equation}
D(X, Y) =
1 - \frac{
\int_0^{12}{
W(z, X) W(z, Y)
}\, dz
}{
\sqrt{\int_0^{12} W(z, X)^2 \, dz}
\sqrt{\int_0^{12} W(z, Y)^2 \, dz}
} \label{eq:spectral_distance}
\end{equation}

\noindent with $W$ as defined in Equation \ref{eq:pc_spec}.
The measure takes values in the interval $[0, 1]$,
where 0 indicates maximal similarity and 1 indicates maximal divergence.

\subsubsection{Harmonicity}

Our harmonicity model is inspired by the template-matching algorithms
of \cite{Milne2013} and \cite{Parncutt1994}.
The model simulates how listeners 
search the auditory spectrum for occurrences of harmonic spectra.
These inferred harmonic spectra are termed 
\textit{virtual pitches}.
High harmonicity corresponds to a strong virtual pitch percept.

Our model differs from previous models in two ways.
First, it uses a pitch-class representation, not a pitch representation.
This makes it voicing-invariant and hence more suitable for modelling pitch-class sets.
Second, 
it takes into account the strength of all virtual
pitches in the spectrum, not just the 
strongest virtual pitch. 

The model works as follows.
The \textit{virtual pitch-class spectrum}
$Q$ defines the spectral similarity 
of the pitch-class set $X$
to a harmonic complex tone with pitch class $p_c$:


\begin{equation}
Q(p_c, X) = D(p_c, X)
\end{equation}

\noindent with $D$ as defined in
Equation \ref{eq:spectral_distance}.
Normalising $Q$ to unit mass produces $Q'$:

\begin{equation}
Q'(p_c, X) =
\frac{Q(p_c, X)}{\int_0^{12}{Q(y, X)} \, dy} .
\end{equation}

\noindent
Previous models compute harmonicity by
taking the peak of this spectrum.
In our experience this works for small chords
but not for larger chords,
where several virtual pitches need
to be accounted for.
We therefore instead compute a
spectral peakiness measure.
Several such measures are possible,
but here we use Kullback-Leibler divergence
from a uniform distribution.
$H(X)$, the harmonicity of a pitch-class set $X$,
can therefore be written as follows:

\begin{equation}
H(X) =
\int_{0}^{12}{
Q'(y, X)
\log_2{\left(12 \, Q'(y, X)\right)}
} \, dy.
\end{equation}

Harmonicity has a large negative correlation
with the number of notes in a chord.
Some correlation is expected, but not to this degree:
the harmonicity model considers a tritone (the least consonant two-note chord)
to be more consonant than a major triad 
(the most consonant three-note chord).
We therefore separate the two phenomena by 
adding a `chord size' feature, corresponding to the number of notes in a given chord, and
rescaling harmonicity to zero mean and unit
variance across all chords
with a given chord size.



\subsubsection{Roughness}

Roughness has traditionally been considered to be an important
contributor to sensory consonance,
though some recent research disputes its importance \cite{McDermott2010}.
We originally planned to include roughness in our model,
but then discovered that the phenomenon is highly sensitive to
chord voicing.
%
%
%
%
Since the voicing of a pitch-class set is undefined,
its roughness is therefore unpredictable.
Roughness is therefore not modelled in the present study.

\subsubsection{Voice-Leading Distance}

A \textit{voice leading} connects
the individual notes in two pitch-class sets
to form simultaneous melodies \cite{Tymoczko2006}.
Pitch-class sets of different sizes can be connected
by allowing pitch classes to participate in multiple melodies.
\textit{Voice-leading distance} is an aggregate measure
of the resulting melodic distance.
We operationalise voice-leading distance using \cite{Tymoczko2006}'s geometric model.

Consider two pitch-class sets
$X = \left\{ x_1, x_2, \ldots, x_{m} \right\}$
and
$Y = \left\{ y_1, y_2, \ldots, y_{n} \right\}$.
A voice-leading between $X$ and $Y$ can be written
$A \rightarrow B$
where
$A = \left( a_1, a_2, \ldots, a_{N} \right)$,
$B = \left( b_1, b_2, \ldots, b_{N} \right)$,
and the following holds:
if $x \in X$ then $x \in A$,
if $y \in Y$ then $y \in B$,
if $a \in A$ then $a \in X$,
if $b \in B$ then $b \in Y$,
and $n \leq N$.



The distance of the voice leading $A \rightarrow B$
is denoted $V(A, B)$ and uses the \textit{taxicab} norm:

\begin{equation}
V(A, B) = \sum_{i=1}^N {d(a_i, b_i)}
\end{equation}

\noindent
with $d(a_i, b_i)$ as defined in Equation \ref{eq:pitch_class_distance}.

The voice-leading distance between pitch-class sets $X, Y$
is then defined as the smallest value of $V(A, B)$
for all legal $A, B$.
This minimal distance can be efficiently computed 
using the algorithm described in \cite{Tymoczko2006}.

\subsubsection{Summary}

This section defined three sensory consonance features.
These included one instantaneous measure
(harmonicity)
and two sequential measures 
(spectral distance, voice-leading distance).
Harmonicity correlated strongly with chord size,
which could have confounded our analyses. 
We therefore controlled for chord size by normalising harmonicity for each chord size and including chord size as a feature. 

\subsection{Statistical Model}

\subsubsection{Overview}


The statistical model is \textit{generative},
defining a probability distribution over chord sequences
(e.g. \cite{Hedges2016a, Paiement2005, Rohrmeier2012e}).
It is \textit{feature-based},
using features of the chord and its context 
to predict chord probabilities
(e.g. \cite{Hedges2016a}).
It is \textit{energy-based},
defining scalar energies for each feature configuration
which are then transformed and normalised to produce the
final probability distribution
(e.g. \cite{Boulanger-Lewandowski2012,Hadjeres2016,Pickens2005}).
It is \textit{exponential-family} in that the energy function
is a linear function of the feature vector
(e.g. \cite{Hadjeres2016,Pickens2005}).
Informally, the model might be said to generalise
linear regression to symbolic sequences.

\subsubsection{Form}
Let $\mathcal{A}$ denote the set of all possible chords,
and let \(e_0^n\) denote a
chord sequence of length \(n\),
where $e_0$ is always a generic start symbol.
Let $e_i \in \mathcal{A}$ denote the $i$th chord
and \(e_i^j\) the subsequence \((e_i, e_{i+1}, \ldots, e_j)\).
Let $\mathbf{w}$ be the 
weight vector that parametrises the model.

The probability of a chord sequence is factorised into
a chain of conditional chord probabilities.

\begin{equation}
P\left(e_0^n\, \vert \, \textbf{w}\right) =
\prod_{i=1}^n{P\left(e_{i}\, \vert \, e_0^{i-1}, \textbf{w}\right)}
\end{equation}

\noindent
These are given energy-based expressions:

\begin{equation}
P\left(e_{i}\, \vert \, e_0^{i-1}, \textbf{w} \right) =
\frac{\exp{(- E(e_0^{i-1}, e_{i}, \textbf{w}))}}{Z(e_0^{i-1}, \textbf{w})}
\end{equation}

\noindent
where $E$ is the \textit{energy function}
and $Z$ is the \textit{partition function}.
$Z$ normalises the probability distribution to unit mass:

\begin{equation} \label{eq:partition_function}
Z(e_0^{i-1}, \textbf{w}) =
\sum_{x \in \mathcal{A}}{\exp{(- E(e_0^{i-1}, x}, \textbf{w}))} .
\end{equation}

High $E$ corresponds to low probability.
$E$ is defined as a sum of
\textit{feature functions}, $f_j$,
weighted by $-\textbf{w}$:

\begin{equation} \label{eq:energy_function}
E(e_0^{i-1}, x, \textbf{w}) = - \sum_{j=1}^m{f_j(e_0^{i-1} :: x) w_j}
\end{equation}

\noindent where $w_j$ is the $j$th component of $\mathbf{w}$,
$m$ is the dimensionality of $\mathbf{w}$, 
equalling the number of feature functions $f_j$,
and $e_0^{i-1} :: x$ is the concatenation of
$e_0^{i-1}$ and $x \in \mathcal{A}$.

Feature functions measure a
property of the last element of a sequence.
Our feature functions are 
chord size, harmonicity,
spectral distance, and voice-leading distance.
Chord size and harmonicity are context-independent,
whereas spectral and voice-leading distance relate 
the last chord to the penultimate chord.
When the penultimate chord is undefined,
mean values are imputed for spectral and voice-leading distance,
with the mean computed over all possible chord transitions.




%
%



\subsubsection{Estimation}

The model is parametrised by the weight vector $\textbf{w}$.
This weight vector is optimised using maximum-likelihood estimation
on a corpus of sequences, as follows.

Let $e_{0,k}^{n_k}$ denote the $k$th sequence from a corpus of size $N$,
where $n_k$ is the sequence's length.
The negative log-likelihood of the weight vector $\textbf{w}$ with respect to the corpus is then

\begin{align} \label{eq:cost}
C(\textbf{w}) &=
- \sum_{k=1}^N {
\sum_{i=1}^{n_k} {
\log {P(e_{i,k} \vert e_{0,k}^{i-1}, \textbf{w})}
}} .
\end{align}

\noindent After some algebra, the gradient can be written

\begin{equation} \label{eq:cost_gradient}
\frac{dC}{d\textbf{w}} =
\sum_{k=1}^N {
\sum_{i=1}^{n_k} {
\frac{Z'(e_{0,k}^{i-1}, \textbf{w})}
{Z(e_{0,k}^{i-1}, \textbf{w})}
- \textbf{f}(e_{0,k}^i)
}}
\end{equation}

\noindent where

\begin{equation}
Z'(e_{0,k}^{i-1}, \textbf{w}) =
\sum_{x \in \mathcal{A}}{
\textbf{f} (e_{0,k}^{i-1} :: x)
\exp{(- E(e_{0,k}^{i-1}, x}, \textbf{w}))
}
\end{equation}

\noindent and $\textbf{f}$ is the vector of feature functions.
This expression can be plugged into a generic optimiser
to find a weight vector minimising the negative log-likelihood.
The present work used the BFGS optimiser \cite{Sun2006}.

\subsubsection{Feature Importance}

This section introduces three complementary 
feature importance measures.
These are \textit{weight}, \textit{explained entropy},
and \textit{unique explained entropy}.

\textit{Weight} describes a feature's 
relationship to chord probability.
The weight for a feature function $f_j$ is the parameter $w_j$,
corresponding to (minus)
the change in the energy function $E$
in response to a one-unit change in the feature function $f_j$
(Equation \ref{eq:energy_function}).
Weight is a signed feature importance measure:
the sign dictates whether the model prefers high (positive weight)
or low (negative weight) feature values,
and the magnitude dictates the strength of preference.
To aid weight comparability between features,
feature functions are scaled to unit variance
over the set of all possible chord transitions.



Dividing the cost function (Equation \ref{eq:cost})
by the number of chords in the corpus ($\sum_{k=1}^N n_k$)
gives an estimate of \textit{cross entropy} in units of \textit{nats}.
Cross entropy measures
chord-wise unpredictability with respect to a given model.
From it we define two further measures:
\textit{explained entropy} and \textit{unique explained entropy}.

\textit{Explained entropy} for a feature $f_j$
is computed by comparing cross entropy estimates for two models:
a model trained using feature $f_j$
and a null model trained with no features.
Explained entropy is the difference between the two cross entropies.
Higher values indicate that the feature explains a lot of structure
in the corpus.

\textit{Unique explained entropy} for a feature $f_j$
is the amount that cross entropy changes when
feature $f_j$ is removed from the full feature set.
It measures the unique explanatory power of a feature
while controlling for other features.


\subsubsection{Related Work}


The literature contains several alternative approaches
for feature-based modelling of chord sequences.
One is the \textit{multiple viewpoint method}
\cite{Harrison2017a,Hedges2016a}.
However, this method is specialised for
discrete features, not the continuous features required for consonance modelling.
A second alternative is the \textit{maximum-entropy} approach
of \cite{Hadjeres2016,Pickens2005}.
This approach has some formal similarities with the present work,
but its binary feature functions are incompatible
with our continuous features.
A third possibility is the \textit{feature-based dynamic networks}
of \cite{Rohrmeier2012e};
however, these networks would need substantial modification 
to represent the kind of feature dependencies required here.



\subsection{Corpora}

Our corpora represent three musical genres:
classical music (1,022 movements/pieces),
popular music (739 pieces),
and jazz music (1,186 pieces).
The classical corpus was compiled from \textit{KernScores} \cite{Sapp2005},
including ensemble music and keyboard music
from several several major composers
of common-practice tonal music (Bach, Haydn, Mozart, Beethoven, Chopin).
Chord labels were obtained using the algorithm of \cite{Pardo2002}
with an expanded chord dictionary,
and with segment boundaries co-located with metrical beat locations as estimated from time signatures.
Chord inversions were identified as the lowest-pitch chord tone
in the harmonic segment being analysed.
The popular and jazz corpora corresponded to publicly available datasets:
the McGill \textit \textit{Billboard} corpus \cite{Burgoyne2011}
and the \textit{iRB} corpus \cite{Broze2013}.


\subsection{Efficiency}


Computation can be reduced by identifying repeated terms
in the cost and cost gradient (Equations \ref{eq:cost}, \ref{eq:cost_gradient}).
These repeated terms only need to be evaluated once.
Our feature functions never look back further than the previous chord,
and they are invariant to chord transposition;
this means that repeated terms occur whenever a 
chord pair is repeated at some transposition.
Collapsing over these repetitions reduces computation 
by a factor of 20--100 for our corpora.

\subsection{Numeric Integration}
The features related to pitch-class spectra all use integration.
These integrals are numerically approximated using the rectangle rule with 
1,200 subintervals, after \cite{Milne2011}.

\subsection{Software}

The statistical model was implemented in R and C++;
source code is available from the authors on request.


\section{Results}\label{sec:results}

\subsection{Corpus level}

Figure \ref{fig:feature_importance_corpus}
plots feature importances for the three consonance measures: harmonicity (normalised by chord size), spectral distance, and voice-leading distance.
Analyses are split by musical corpus, and
confidence intervals are calculated using
nonparametric bootstrapping \cite{Efron1993}.

\subsubsection{Importance by Feature}

All the consonance features contribute to harmonic structure in some way.
The order of feature importance is fairly consistent between
genres and importance measures.
Broadly speaking,
voice-leading distance is most important,
followed by harmonicity,
then spectral distance.


\subsubsection{Importance by Corpus}

Harmonicity is particularly important for popular music,
less so for classical,
and least for jazz.
Spectral distance is most important for classical music,
less so for popular, and unimportant for jazz.

The relative importance of voice-leading distance depends on the measure used:
it scores highly on explained entropy,
but less on weight and unique explained entropy.
This may be because voice-leading distance and 
chord size capture some common information:
moving from a small chord to a large chord typically
involves a large voice-leading distance.
If we wish to assess the unique effect of voice-leading distance,
we can look at weight and unique explained entropy:
these measures tell us that
voice-leading distance is most important for jazz music,
less for classical music,
and least for popular music.

\subsubsection{Signs of Weights}

The sign of a feature weight determines whether the model
prefers positive or negative values of the feature.
The observed feature signs are all consistent with theory.
Harmonicity has a positive weight for all genres,
indicating that harmonicity is universally promoted.
Spectral distance and voice-leading distance
both have negative weights,
indicating preference for lower values of these features.


%

\subsection{Composition Level}

\begin{figure*}
 \includegraphics[width=\textwidth]{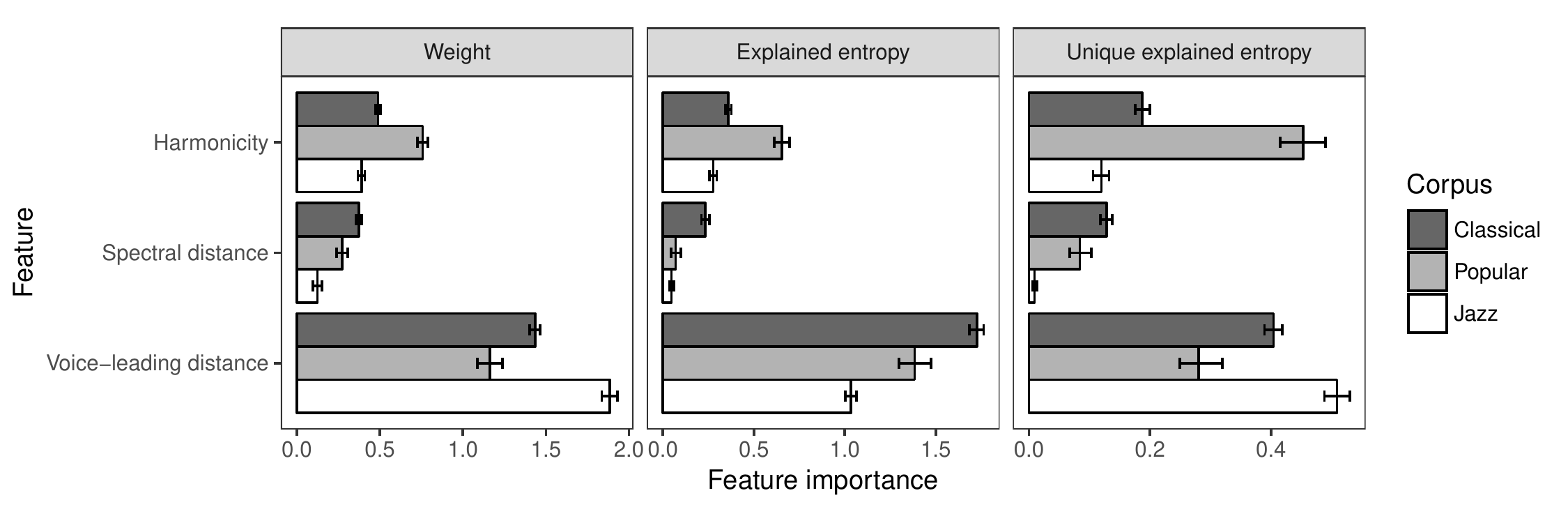}
 \caption{Measures of feature importance as a function of musical corpus. These measures are calculated from statistical models trained on the corpus level. Error bars represent 99\% confidence intervals estimated by nonparametric bootstrapping
\cite{Efron1993}.
Signs of feature weights are reversed for spectral distance and voice-leading distance,
so that positive weights correspond to consonance maximisation.
}
 \label{fig:feature_importance_corpus}
\end{figure*}

\begin{figure*}
 \includegraphics[width=\textwidth]{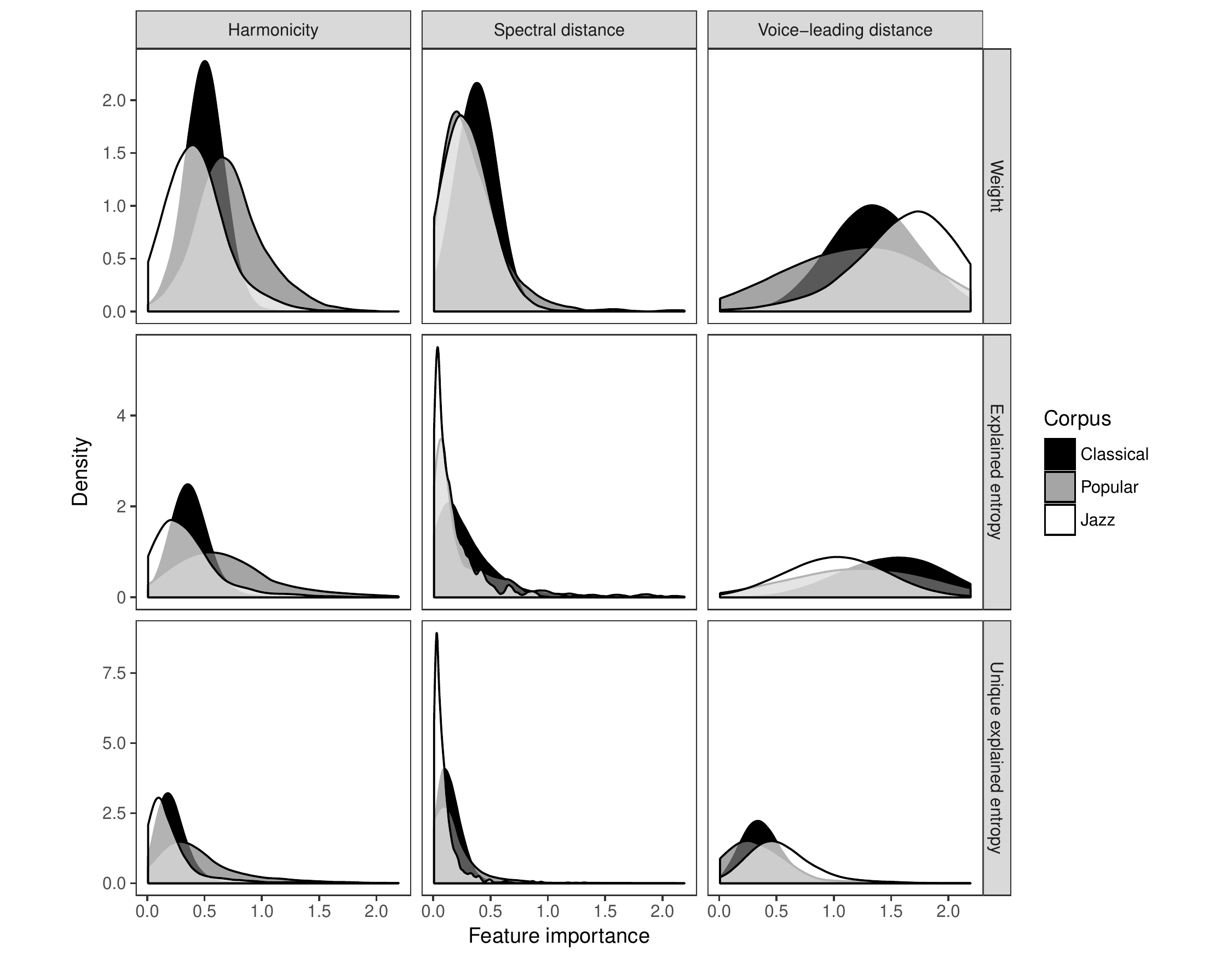}
 \caption{Distributions of feature importance measures as calculated for individual compositions within the three corpora. Distributions are represented by Epanechnikov kernel density functions.
Signs of feature weights are reversed for spectral distance and voice-leading distance,
so that positive weights correspond to consonance maximisation.
}
 \label{fig:feature_importance_composition}
\end{figure*}

We also explored the application of these techniques 
to individual compositions
(Figure \ref{fig:feature_importance_composition}).
While the composition-level analyses reflect the same trends
as the corpus-level analyses
(Figure \ref{fig:feature_importance_corpus}),
they also reveal substantial overlap between the corpora.
We assessed the extent of this overlap by training a generic
machine-learning classifier to predict genre
from the complete set of feature importance measures.
%
%
%
Our classifier was a random forest model 
trained using the randomForest package in R \cite{Liaw2002},
with 2,000 trees and four variables sampled at each split.
Performance was assessed using 10-fold cross-validation repeated
and averaged over 10 runs,
resulting in a classification accuracy of 86\%
and a kappa statistic of .79.
This indicates that genre differences in sensory consonance are moderately salient, even at the composition level.



%


\section{Conclusion}\label{sec:conclusion}

This paper introduces new methods for testing relationships between sensory consonance and Western harmony.
The methods centre on a new statistical model that predicts symbolic sequences using continuous features.
We demonstrate these methods through application to three corpora 
representing classical, popular, and jazz music.


The results strongly support theoretical relationships between 
sensory consonance and harmonic structure. 
The three aspects of sensory consonance tested 
-- harmonicity, spectral distance, and voice-leading distance --
all predicted harmonic movement.
Not all aspects were equally important, however.
Spectral distance performed poorly, particularly in jazz.
This is interesting given the high importance
attributed to spectral distance in recent psychological literature
\cite{Bigand2014,Collins2014,Milne2016a}.
Harmonicity performed well in popular music, but less so in classical and jazz.
In contrast, voice-leading distance performed consistently well.

%
%
%


The corpus analyses deserve further development.
It would be worth probing the true universality of sensory consonance
by exploring a broader range of styles
and using more refined stylistic categories,
possibly at the level of the composer.
The validity of the classical analyses could also be 
improved through more principled sampling \cite{London2013}
and manual chord-labelling \cite{Jacoby2015}.

The three feature importance measures provide useful complementary perspectives, 
but it is unnecessary to plot each one every time.
In future we recommend inspecting the weights
to check whether a feature is promoted or avoided,
but then plotting just unique explained entropy. 
Unique explained entropy is preferable to weight because its units are well-defined,
and preferable to explained entropy because it controls for other features, 
thereby providing a better handle on causality.

We focused on interpreting the statistical model through feature importance measures, 
but an alternative strategy would be to use the model to generate chord sequences for subjective evaluation.
This route lacks the objectivity of feature-importance analysis,
but it would give a uniquely intuitive perspective on what the model has learned.

The modelling techniques could be developed further.
An important limitation of the current model is the linearity 
of the energy function, which restricts it to 
monotonic feature effects.
A polynomial energy function would address this problem.
It would also be interesting to develop the psychological features further,
perhaps adding echoic memory to the spectral distance measure 
\cite{Leman2000},
and introducing an octave-generalised roughness measure.

Despite these limitations, 
we believe that the current results have important implications for our understanding of Western tonal harmony. 
In particular, the results imply that voice-leading efficiency is a better candidate for a harmonic universal than spectral similarity.
This result is important for music psychology,
where voice-leading efficiency is relatively underemphasised compared to harmonicity and spectral similarity (though see \cite{Parncutt1989,Bigand1996,Tymoczko2006}).
Future psychological work may wish to re-examine the role of voice-leading efficiency in harmony perception.




\pagebreak

\section{Acknowledgements}

The authors would like to thank Emmanouil Benetos and Matthew Purver for useful feedback and advice regarding this project. PH is supported by a doctoral studentship from the EPSRC and AHRC Centre for Doctoral Training in Media and Arts Technology (EP/L01632X/1).

\bibliography{bib}
%
%
%

\end{document}